\documentclass[12pt]{article}
\usepackage{amsmath2000,amssymb}
\usepackage{epsfig}

\newcommand\be{\begin{equation}}
\newcommand\ee{\end{equation}}
\newcommand{\bea}{\begin{array}}
\newcommand\ea{\end{array}}
\newcommand\beqa{\begin{eqnarray}}
\newcommand\eeqa{\end{eqnarray}}
\newcommand\bean{\begin{eqnarray*}}
\newcommand\eean{\end{eqnarray*}}

\begin{document}
\baselineskip=15pt
\begin{flushright}
SU-4240-758
\end{flushright}

\vskip 1cm

\begin{center}
{\Large\bf
Quantum Spacetimes in the Year 1}
\footnote{Talk given at the SUNY Institute of Technology at Utica/Rome
Fall 2001 Conference on Theoretical High Energy Physics and  at the $2^{nd}$
Winter Institute on Foundations of Quantum Theory and Quantum Optics 
Quantum Information Process ( January 2-11, 2002), S.N.Bose National Centre 
for Basic Sciences, Kolkata.}

\vskip .8cm

A.P.Balachandran 
\footnote{Electronic address: {\tt bal@phy.syr.edu.}}

\vskip .8cm
{ \it
Physics Department,Syracuse  University , Syracuse, New York 13244-1130.}
\\

\end{center}

							    \vskip 1cm

\vskip 1cm	

\begin{center}
{\bf  Abstract}\\
\end{center}

{\small

We review certain emergent notions on the nature of spacetime from 
noncommutative geometry and their radical implications. These ideas of 
spacetime are suggested from developments in fuzzy physics, string theory,
and deformation quantisation. The review focuses on the ideas coming from fuzzy
physics. We find models of quantum spacetime like fuzzy $S^4$ on which 
states cannot be localised, but which fluctuate
into other manifolds like $ CP^3$ .
New uncertainty principles concerning such lack of localisability on quantum 
spacetimes are formulated.Such investigations show the possibility of 
formulating and answering questions like the probabilty of finding a 
point of a quantum manifold in a state localised on another one. Additional 
striking possibilities indicated by these developments is the ( generic )
failure of $CPT$ theorem and the conventional spin-statistics connection.
They even suggest that  Planck's `` constant '' may not be a
constant, but an operator which does not commute with all observables.
All these novel possibilities arise within the rules of conventional
quantum physics,and with no serious input from gravity physics.

}

\sf

\newpage

\setcounter{footnote}{0}

\section{Spacetime in Quantum Physics}

The point of departure from classical to quantum physics is the 
algebra $\mathcal{F}(T^*Q)$ of functions on the classical phase 
space $T^*Q$. According to Dirac, quantisation can be achieved 
by replacing a function $f$ in this algebra by an operator 
$\hat f$ and equating $i\hbar$ times the Poisson bracket 
between functions to the commutator between the corresponding 
operators. 

In classical physics, the functions $f$ commute, so 
$\mathcal{F}(T^*Q)$ is a commutative algebra. 

But the corresponding quantum algebra $\hat{ \mathcal{F}}$ is 
not commutative. Dynamics is on $\hat{\mathcal{F}}$. So quantum 
physics is {\it noncommutative dynamics}. 

A particular aspect of this dynamics is {\it fuzzy phase space} 
where we cannot localise points, and which has an attendent 
effective ultraviolet cutoff: The number of states in a phase 
space volume $V$ is infinite in classical physics and 
$V/{\hbar^{2d}}$ in quantum physics when the phase space is of 
dimension $2d$. The emergence of this cutoff from quantisation 
is of particular importance for the program of fuzzy physics 
\cite{Madore1}. 

This brings us to the focus of our talk. In quantum physics, 
the commutative algebra of functions on phase space is deformed 
to a noncommutative algebra, leading to a ``noncommutative 
phase space''. Such deformations, characteristic of quantum 
theory, are now appearing in different approaches to 
fundamental physics. The talk will focus on a few such selected 
approaches and their implications. 

Before proceeding further, let us mention a few of these lines 
of thought leading to noncommutative geometry: 1) 
Noncommutative geometry has made its appearance as a method for 
regularising quantum field theory (qft) and in studies of 
deformation quantisation. This talk will more or less base 
itself on these aspects. 2) It has turned up in string physics 
as quantised D-branes. 3) Certain approaches to canonical 
gravity \cite {Paulo} have also used noncommutative geometry 
with great effectiveness. 

\section{Fuzzy physics and quantum field theory}

In what follows, we will focus on fuzzy physics both as a means 
to regularise quantum field theories and in their relation to 
quantum spacetimes. 

But as mentioned already, we will not talk about how they 
emerge from string physics. 

\section{Fuzzy spacetime as regulator}

The original ideas for using fuzzy spaces to regulate qft's are 
due to Madore \cite{Madore1}. They concern quantising 
underlying spacetime itself, making it into a {\it fuzzy 
spacetime}. 

We know since Planck and Bose that quantisation introduces a 
short distance cut-off, changing the number of states in a 
phase space volume $V$ from $\infty$ to $V$$/$${\hbar}^{2d}$. 

Now qft's on a manifold require regularisation. The usual 
nonperturbative regularisation involves lattice qft's. The use 
of fuzzy spacetimes can be another. The latter has important 
advantages like maintaining symmetries and avoiding fermion 
doubling. The particular approach reported here involves many 
colleagues. Our representative papers are \cite{Bal1,Bal2} and 
the work of Vaidya, Dolan et al. and Lopez et al. in 
\cite{Sachin1}. Related or overlapping work is due to \cite{Shahin1}
and \cite{Ram}.

We do Euclidean qft's. Quantising $S^4$ would be of the 
greatest physical interest. But for now, we will focus on the 
simpler case of $S^2$. 

\section{Fuzzy ${\mathbb C}^2$, $S^3$ and $S^2$}

\subsection{Relation between ${\mathbb C}^2$, $S^3$ and $S^2$} 

Consider ${\mathbb C}^2$ with coordinates $z=({z_1},{z_2})$. We have
\be
S^{3} = <z:\mid z \mid {^2} := \sum \mid z_{i}\mid ^{2}=1>\subset 
{\mathbb C}^{2}
\label{SSS}
\ee
and 
\beqa
S^{2}&=&<\vec{x}=z^{\dagger}{\vec\tau}z,z\in S^3 >,\\
\vec\tau &=&{ Pauli}\quad { matrices,}\\
{\vec{x}.\vec{x}}&=&1.
\label{SS} 
\eeqa
$S^2$ is the quotient of $S^3$ by the $U(1)$ action 
$z\rightarrow{z{e^{i\theta}}}$ (Hopf fibration). 

The group $SU(2)=\{g\}$ acts on ${\mathbb C}^2$, $S^3$ and $S^2$:
\beqa
z\rightarrow gz,
x_i\rightarrow R_{ij}(g)x_j,
R(g)= {Rotation}\quad{ matrix}\quad{for}\quad{g}.
\label{action}
\eeqa

\subsection{The fuzzy ${\mathbb C}^2$}

Now we {\it quantise} ${\mathbb C}^2$ by the replacements 
\beqa
z_{\alpha}\rightarrow a_{\alpha},
\bar{z_{\alpha}}\rightarrow a_{\alpha}^{\dagger}, 
\eeqa  
where $a_{\alpha}$ and $a_{\alpha}^{\dagger}$ are harmonic 
oscillator annihilation and creation operators with the usual 
commutation relations. $SU(2)$ still acts with generators 
\be
L_\alpha =  a^{\dagger}{\sigma}_\alpha a
\ee
which commute with the number operator:
\beqa
N=a^{\dagger} a,
[N,L_\alpha]=0.
\eeqa

\subsection{The fuzzy three-sphere}

Consider
\beqa
S_{\alpha}&=&{a_{\alpha}}\frac{1}{\sqrt{N+1}}, \nonumber \\
S{_{\alpha}}{^{\dagger}}&=&\frac{1}{\sqrt{N+1}}
a_{\alpha}{^\dagger}, \nonumber \\
S{_{\alpha}}{^{\dagger}}S{_{\alpha}}&=&\frac{N}{N+1} \rightarrow 1 \,
{as}\, N\rightarrow \infty,
\nonumber \\
\left[S_{\alpha},S_{\beta}^{\dagger}\right]&\rightarrow& 0\, 
{as}\, N\rightarrow \infty.
\eeqa
So $S_{\alpha}$ are normalised commuting vectors as 
$N\rightarrow\infty$ and the fuzzy three-sphere $S{_F}{^3}$ is 
the algebra generated by $S_{\alpha}$ and 
$S{_{\beta}}{^{\dagger}}$. 

{\it It is important to note that $\frac{1}{(N+1)}$ here plays 
the role of a quantised Planck's constant.} 

\textit{This raises the following important questions:Is it possible that 
Planck's ``constant'' is in reality an operator ? How can one experimentally 
test this possibilty?}

The representation of the $S{_F}{^3}$-algebra on the Fock space 
is irreducible. 

\subsection{The fuzzy two-sphere}

Since $[N,L_\alpha]=0$, we have that
\be
[N,S{^{\dagger}{\sigma}_\alpha}S]=0.
\ee
So we can restrict the algebra of the fuzzy sphere $ S{^2}_{F}$ 
generated by $S{^{\dagger}{\sigma}_\alpha}S$ to the finite-
dimensional vector space spanned by the eigenvectors of $N$ 
with eigenvalue $n$. They are spanned by 
\beqa
|n_1,n_2>&:=&\frac{1}{\sqrt{n_1!n_2!}} 
(a{_1}{^\dagger})^{n_1}
(a{_2}{^\dagger})^{n_2}|0>,
\nonumber\\
{n_1}+{n_2}&=&n,
\label{state}
\eeqa
and carry angular momentum $J{_0}=\frac{n}{2}$. As this 
representation is irreducible, it follows that the fuzzy sphere 
algebra for angular momentum $J_0$ is the algebra of $(2J_0+1)$ 
x $ (2J_0+1)$ matrices $Mat(2J_0+1)$. We denote its elements by 
$M$. It is just the vector space spanned by the tensor 
operators in the angular momentum $n/2$ representation. 

Rotation acts on $M$ by $M \rightarrow gMg^{-1}$. So the fuzzy 
sphere has angular momenta $0,1,...,2J_0$ , cut off at $ 2J_0$. 
The algebra of functions on the two-sphere instead has all 
integral angular momenta upto $\infty$. 

\section{Scalar fields on $S{_F}{^2}$}

Scalar fields are power series in ``coordinates" 
$S{^{\dagger}{\sigma}_\alpha}S$.
So
\be
Scalar\, Field = (2J_0 + 1 )-dimensional\quad matrix.
\ee
A scalar action, such as
\be
\mathcal{A} = \frac{1}{2J_0+1} Tr ([L_i,\phi]{^\dagger}[L_i,\phi] +
\phi^{k})
\ee
can be quantised by functional integral methods. 
Renormalisation studies of such actions have been carried out 
in \cite{Sachin1}. 

Gauge theories can be formulated on fuzzy spaces. For brevity, 
we will not enter into their discussion. 

Summarising, we have the classical descent chain 
\be
\mathbb {C} \rightarrow S^3 \rightarrow S^2.
\ee
It becomes after quantisation
\be
\mathbb{C}_{F}\rightarrow S{_F}{^3}\rightarrow S{_F}{^2}.
\ee
The algebra dimension is $\infty$ for all except $S{_F}{^2}$ for which
it is $ (2J_0 +1)^2$.

\section{On coherent states and star products}

In quantum field theory, we calculate correlation functions like
\be
<\phi (\vec{n}{_1})\phi(\vec{n}{_2})\ldots\phi(\vec{n}{_j})> ,\qquad 
\vec{n}{_j}
\in S^2.
\ee
To compare with such expressions, we have to know how to map 
our operators (finite-dimensional matrices) to functions. This 
is where coherent states and star products prove important. For 
us, they will be particularly important also when discussing 
issues of topology fluctuations. We now explain how star 
products can be defined on fuzzy spaces. 

We start from the infinte dimensional Fock space associated with two
oscillators and introduce the standard coherent states
\be
\mid z;\infty>=e^{z_{\alpha}a_{\alpha}{^\dagger}-\bar z_{\alpha}a_{\alpha}
}|0>
\ee
where $\infty$ has been inserted in the state vector to 
indicate that it is associated with the Fock space. (It is 
omitted from the vacuum stae which will be commom to both the 
Fock space and its subspaces which will appear below.) 

A theorem asserts that the diagonal matrix elements
\be
<z;\infty\mid \hat A\mid z;\infty>=A(z)
\ee
determines the operator $\hat A$ (under suitable conditions on 
$\hat A$): the map $\hat A \rightarrow A$ of operators to 
functions is one-to-one. Both the Moyal and Perelomov star 
products follow from this result as we now indicate. 

The star product $A*B$ of two functions $A$ and $B$ is defined by
\be
A*B(z):=<z;\infty\mid \hat A\hat B\mid z;\infty>.
\ee
(It is not the Moyal star product, but equivalent to it.)

If $\hat{A}$ is $0$ outside the subspace where $N=n$,then
\beqa
\hat{A}&=&P\hat{A}P,\nonumber\\
P&=&\textrm{Projector\, onto\, this\, subspace}.
\eeqa
The explicit expression for $P$ is 
\be
P=\sum_{n_1+n_2=n}{\mid} n_1,n_2><n_1,n_2\mid
\ee
where we have used the definition (\ref{state}).

The diagonal coherent state expectation value of $P\hat{A}P$ is (upto a 
constant in the definition of the state)
\beqa
A(z)&=&<z| \hat{A}|z>, \nonumber \\
|z>&:=&\frac{1}{\sqrt{n!}}\sum_{n_1+n_2=n}(z_\alpha a_{\alpha}{^\dagger})^n|0>.
\eeqa
The group $SU(2)$ with generators $L_i$ acts on these states according to
\beqa
|z>&\rightarrow&|gz>, \nonumber \\
g&\in& SU(2)
\eeqa
preserving $\sum{|z_i|^2}$, so we can set it equal to $1$. This 
normalises these states. 

The fuzzy sphere operators $L_i$ and $\frac{1}{N+1}L_i$ can be  
restricted to $N=n$. We can set them equal to $0$ on the 
subspace $N\neq{n}$, and treat them as the above $\hat{A}$'s. 

In this way, we have a star product on $S{_F}{^2}$. There are explicit 
expressions for this star product\cite{Bal2}.

\section{Fluctuating topologies: A novel space-time uncertainty 
principle} 

We now pass to the 4-sphere $ S^4$ and consider the fuzzy $S^4$, which is 
denoted by $S{_F}{^4}$.( See in this connection\cite{Ram}.) We realise  $S^4$ as a sphere in 
$\mathbb{R}{^5}$:
\be
S^4=<x:=(x_1,x_2,..,x_5):\sum|x_i|^2=1>.
\ee
For quantisation, it is necessary to introduce $\mathbb{C}{^4}$ with 
coordinates $z=(z_1,z_2,..,z_4)$. The unit sphere in $\mathbb{C}{^4}$ is 
$S^7$:
\be
S^7=<z:\sum|z_i|^2=1>.
\ee
We also introduce the five gamma matrices $\gamma_{\lambda}$:
\be
\gamma_{\lambda}=\,standard\,{\gamma}\,matrices.
\ee
Then we can set
\be
x_{\lambda}=z^{\dagger}\gamma_{\lambda}z,\nonumber\\
z\in S^7.
\ee

Now instead of $2$, we have $4$ sets of creation-aniihilation 
operators $a_{\alpha},a_{\alpha}{^\dagger}$ ($\alpha=1,2,3,4$). 
$S{^4}$ will be thought of as a sphere in $\mathbb{R}{^5}$ and 
its quantised fuzzy version will hence be formulated. 

Coherent states appropriate for $S{^4}$ are generalisations of 
those for $S{^2}$: 
\be
|z>:=\frac{1}{\sqrt{n!}}\sum_{n_1+n_2+n_3+n_4=n}
(z_\alpha a_{\alpha}{^\dagger})^n|0>
\ee
where the sum on $\alpha$ is over $4$ values.

The quantised version of $x_{\lambda}$ is $S{^{\dagger}{\sigma}_\alpha}S$
with $S$'s being defined as before. 

A simple calculation shows that 
\be
<z|S{^{\dagger}{\sigma}_\alpha}S|>= (\frac{n}{n+1})\,
z^{\dagger}\gamma_{\lambda}z.
\ee
Or it gives $S{^4}$ as imbedded in $\mathbb{R}{^5}$.

But we now show that the fuzzy $S{_F}{^4}$ emerges only 
approximately, having fluctuations of the order of 
$\frac{1}{n}$ into the six-dimensional fuzzy manifold 
$\mathbb{C}P{^3}{_F}$, the fuzzy version of $\mathbb{C}P{^3}$. 

To demonstrate this, consider the correlation function
\be
<z|S{^{\dagger}{\sigma}_\alpha}S\,S{^{\dagger}{\sigma}_\beta} |z>.
\ee
A short calculation shows that it is
\be
z^{\dagger}\gamma_{\alpha}z \,z^{\dagger}\gamma_{\beta}z + O(\frac{1}{n})
\,z^{\dagger}\gamma_{\alpha\beta}z
\ee
where $\gamma_{\alpha\beta}$ is 
$[\Gamma_{\alpha},\Gamma_{\beta}]$. Now 
$z^{\dagger}\gamma_{\alpha\beta}z$ is {\it not} a function on 
$S^4$. It is a function only on $\mathbb{C}P{^3}$. (It is 
invariant under the phase change of $z:z\rightarrow 
e^{i\theta}z$.) 

By definition,
\be
<z|S{^{\dagger}{\sigma}_\alpha}S\,S{^{\dagger}{\sigma}_\beta} |z>=
(\frac{n}{n+1})z^{\dagger}\gamma_{\alpha}z*(\frac{n}{n+1})
z^{\dagger}\gamma_{\beta}z.
\ee.

A measure of the lack of localisation of the star product on 
$S^4$ is possible to construct, but we will not discuss that 
here. 

{\it The behaviour of $z^{\dagger}\gamma_{\alpha}z$ under star 
product is generic for most manifolds, notable exceptions being 
$\mathbb{C}P{^N}$.} 

Thus we have an algebra of observables, such as that generated 
by $z^{\dagger}\gamma_{\alpha}z$ under *, which is only {\it 
approximately} localised on a manifold $M$. It has fluctuations 
$O(\frac{n}{n+1})$ into another manifold $M' \supset M$ which 
disappear in the classical limit $\frac{1}{n}\rightarrow0$. 

\section{More on quantum topologies}

The fuzzy space $S{_F}{^2}$ has representations in all 
dimensions. 

In contrast the fuzzy space $S{_F}{^4}$, or rather 
$\mathbb{C}P{_F}{^3}$, has representations in symmetric 
products of its four-dimensional representation, namely in 
dimensions $4,10,20,...$. 

So $20 \times 20$ matrices can approximate {\it either} $S{^2}$ 
{\it or} $\mathbb{C}P{^3}$. 

So what we see in these matrices depends on the operators we 
examine \cite{Madore2}. If we work with $S^2$-coherent states 
and operators appropriate for them, then they will approximate 
$S^2$. But if instead we work with states and operators 
appropriate for $\mathbb{C}P{_F}{^3}$, then these matrices will 
approximate that manifold. 

In high dimensions, such approximate manifolds proliferate.

We can answer questions like: the probabilty of finding a 
$\mathbb{C}P{_F}{^3}$- localised state like 
\beqa
\rho &=& |z><z|, \nonumber \\
z &\in& \mathbb{C}P{^3}.
\eeqa
($|z>$ being the coherent state for $\mathbb{C}P{_F}{^3}$) in a 
$\mathbb{C}P{_F}{^1}$-localised state 
\beqa
\tilde{\rho}&=&|\tilde{z}><\tilde{z}|, \nonumber \\
\tilde{z}&\in&\mathbb{C}P{^1}, \, that\, is \, S^2\,,
\eeqa
($|\tilde{z}>$ being the coherent state for 
$\mathbb{C}P{_F}{^1}$. The answer is 
\be
|Tr{\rho}^\dagger \tilde{\rho}|^2.
\ee

{\it We can see from this discussion that questions with 
fantasy about quantum spacetimes become accessible in fuzzy 
physics.} 

\section{Causality and $CPT$ violation}

Fuzzy models can be formulated for spacetimes with Minkowsky 
metric, not just for Euclidean spacetimes like $S^2$. They are 
natural models to quantise time, preserving many symmetries. 

A popular model for quantising Minkowsky spacetime which has 
been carefully studied by Doplicher, Fredenhagen and Roberts 
\cite{Doplicher} and others \cite{Peter,Shahin2} is governed by the 
commutation relation 
\beqa
[x^\mu,x^\nu]&=&i\theta^{\mu,\nu},\nonumber\\
\theta^{\mu,\nu}&=& -\theta^{\nu,\mu}=a\,real\, constant.
\eeqa

With spacetime fuzzy, the meaning of ``spacelike separation'' 
loses exact meaning and leads to causality and hence generically to $CPT$ 
violation\cite{Shahin2}. 

A way to understand this is as follows. In $1+1$-dimensions, we can set
\be
\theta^{\mu,\nu}= \theta \epsilon^{\mu,\nu}
\ee
where $\epsilon^{\mu,\nu}$ is the Levi-Civita symbol with $\epsilon^{0,1}
=1$. Then \cite{Peter} the fuzzy version of a free field of mass $M$ has
the representation
\be
\Phi=\int \psi(k)e^{ik_0 x_0}e^{ik_1x_1}
\ee
where
\beqa
k^{2}_{0}-k^{2}_{1}&=&M^2, \nonumber \\
\psi(k)&=&\delta(k^2-M^2)a(k),\,k_0 > 0, \nonumber \\
       &=&\delta(k^2-M^2)a(-k_0,k_1)^{\dagger},\,k_0 < 0,
\eeqa
$a(k)$ and $a(k)^{\dagger}$ being the usual annihilation and 
creation operators. 

We can regard $(x_0 + i x_1)/\sqrt{2\theta}$ and its adjoint as 
annihilation and creation operators. If $|z>$ denotes the 
corresponding coherent state, then the square of the fuzzy 
field localised approximately at $(Re\,z$,$Im\,z)$ is 
\be
<z|(\Phi)^2|z> := (\bar{\Phi})^2(z). 
\ee
One checks that when $z$ and $z'$ correspond to spacelike points and 
$\alpha$ and $\beta$ are generic states,
\be
<\alpha|[\bar\Phi)^2(z),(\bar\Phi)^2(z')]|\beta>\approx e^{-|z-z'|^2}.
\ee

Such causality violations are {\it generic} in such models. A 
fundamental question, inspired by these models then is: {\it 
How can we test them?} \cite{Amelino}. We can in principle do 
so using forward dispersion relations, but they are expected to 
be small and it is not clear if these dispersion relations can 
be evaluated with sufficient accuracy to establish significant 
limitations on possible causality violation. 

\section{Winding up}

The following broad observations are suggested by the preceding 
discussions: 

- Models of spacetime based on noncommutative geometry are 
suggested by quantum physics itself, string physics and 
attempts to regularise quantum field theories. 

- They lead to strikingly novel spacetime models.

- But lacking contact with experiments, these models for now 
remain metaphysical models, just as quantum gravity and string 
models. 

\section*{Acknowledgments}
 
The new ideas reported here come from the joint efforts of a 
group of us, some of our representative papers being 
\cite{Bal1,Bal2} and the work of Vaidya, Dolan et al. and Lopez 
et al. in \cite{Sachin1}. This work has been supported by DOE 
under contract number DE-FG02-85ER40231 and by the NSF-CONAcYT 
grant INT-9908763 which facilitated several mutual visits by 
the US and Mexican participants.

\end{document}